# Electrical control of deep NV centers in diamond by means of sub-superficial graphitic micro-electrodes

J. Forneris[1,2,3,]*, S. Ditalia Tchernij[2,1], A. Tengattini[2,1,3], E. Enrico[4], V. Grilj[5],

N. Skukan[5], G. Amato[4], L. Boarino[4], M. Jakšić[5], P. Olivero[2,1,3]

[1]*Istituto Nazionale di Fisica Nucleare (INFN), Sez. Torino, via P. Giuria 1, 10125, Torino, Italy*
[2]*Physics Department and "NIS " Inter-departmental Centre - University of Torino; via P. Giuria 1, 10125, Torino, Italy*
[3]*Consorzio Nazionale Interuniversitario per le Scienze Fisiche della Materia (CNISM), Sez. Torino, Torino, Italy*
[4]*Istituto Nazionale di Ricerca Metrologica (INRiM); Strada delle Cacce 91, 10135 Torino, Italy*
[5]*Ruđer Bošković Institute, Bijenicka 54, P.O. Box 180, 10002 Zagreb, Croatia*

**ABSTRACT**

The control of the charge state of nitrogen-vacancy (NV) centers in diamond is of primary importance for the stabilization of their quantum-optical properties, in applications ranging from quantum sensing to quantum computing. In this work buried current-injecting graphitic micro-electrodes were fabricated in bulk diamond by means of a 6 MeV $C^{3+}$ scanning micro-beam. The electrodes were exploited to control the variation in the relative population of the negative ($NV^-$) and neutral ($NV^0$) charge states of sub-superficial NV centers located in the inter-electrode gap regions. Photoluminescence spectra exhibited an electrically-induced increase up to 40% in the $NV^-$ population at the expense of the $NV^0$ charge state, with a linear dependence from the injected current at applied biases smaller than 250V, and was interpreted as the result of electron trapping at NV sites. An abrupt current increase at ~300V bias resulted in a strong electroluminescence from the NV0 centers, in addition to two spectrally sharp emission lines at 563.5 nm and 580 nm, not visible under optical excitation and attributed to self-interstitial defects. These results disclose new possibilities in the electrical control of the charge state of NV centers located in the diamond bulk, which are characterized by longer spin coherence times.

---

* Corresponding author:  Tel +39 011 6707306,         forneris@to.infn.it,

# 1 INTRODUCTION

The nitrogen-vacancy complex in diamond (NV center) emerged in the last decade as a prominent solid-state quantum system operating at room temperature, and has been exploited for innovative applications in quantum computing and quantum sensing [1-6]. Most of these applications heavily rely on the initialization, manipulation and readout of the spin state of the negatively charged (NV$^-$) state of the defect. However, the NV center is affected by uncontrolled blinking to the neutral charge state (NV$^0$) [7], thus preventing its efficient manipulation [8].

In recent years, several works addressed the control of the charge state of the NV center by different approaches. The chemical stabilization of NV$^-$/NV$^0$ charge state was investigated in shallow centers and nanodiamonds, by means of oxygen [9-11], fluorine [12] and hydrogen [13] surface termination. The NV charge state stability has been investigated as a function of laser power and wavelength excitation [7,14,15]. Furthermore, the electrical control was adopted to stabilize the negative charge state of shallow NV centers in a hydrogen-terminated sample [16,17], and to deterministically switch NV$^-$ centers to the neutral charge state through hole injection in p-i-n devices [18,19].

The charge state control of the NV$^-$ center by means of electrical pulse sequences is particularly appealing, since it would enable the development of integrated spintronic devices [18]. Despite the recent progresses, the current studies were mainly limited to surface (i.e. <14 nm) defects [16,17] or to the stabilization of the NV$^0$ charge state [18,19], with the exception of the negative charge state stabilization in p-i-n devices in a sandwich configuration [20]. On the other hand, the possibility of controlling deep NV$^-$ centers in diamond using flexible electrode geometries discloses appealing opportunities in both quantum metrology and quantum information processing, since deep sub-superficial NV$^-$ centers are characterized by significantly longer spin coherence times due to the weaker interaction with surface defect states [21].

* Corresponding author: jacopo.forneris@unito.it

In this work, the electrical control of the charge state in NV ensembles is investigated in a single-crystal diamond substrate structured with sub-superficial graphitic micro-electrodes fabricated by means of MeV ion beam lithography [22]. The employed fabrication technique allows to define arbitrary electrode geometries with micrometric resolution in the diamond bulk (i.e. up to several micrometers below the sample surface) by exploiting the radiation-induced graphitization of the material occurring at the end of the MeV ion penetration range, and has already been successfully adopted to realize different integrated devices in diamond, such as bolometers [23,24], particle detectors [25], cellular biosensors [26,27] and IR emitters [28]. More pertinently to this work, sub-superficial graphitic electrodes were previously employed to stimulate electroluminescence from diamond color centers, both in multi-photon [29] and single-photon [30] emission regimes. Furthermore, the exploitation of the ion fabrication technique enabled to investigate the charge state conversion process by means of non-rectifying ohmic electrodes, thus taking advantage on the related charge injection mechanisms.

## 2 METHODS

### 2.1 Device fabrication

The employed sample was a 3×3×0.3 mm$^3$ single-crystal diamond substrate grown by ElementSix by CVD technique. This "optical grade" type IIa sample is characterized by nominal concentrations of substitutional nitrogen and boron of <1 ppm and <0.05 ppm, respectively. Two sub-superficial graphitic micro-electrodes were fabricated in the sample bulk by raster-scanning a focused ∅~5 μm 6 MeV $C^{3+}$ beam along linear paths. The ion fluence (~$4\times10^{16}$ cm$^{-2}$) was chosen to overcome the graphitization threshold [31] at the end of the ions range, i.e. at a depth of ~2.7 μm below the sample surface, as shown in **Figure 1a**. The sample was subsequently annealed at 1000 °C in vacuum for 2 hours to convert the amorphized layer to a graphitic phase. This thermal process was concurrently exploited to induce the formation of a high-density ensemble of NV centers in the gap region comprised between the graphitic channels, due to the aggregation of native nitrogen atoms and the vacancies created by stray ions in the fabrication process. Subsequently, the sample underwent a 30 min oxygen plasma treatment (20 sccm $O_2$ flux, 30 W RF power, pressure $2.5\times10^{-2}$ mbar) with the purpose of removing any residual surface conductivity associated with possible graphitization or contamination occurring during the thermal annealing. In addition, the plasma treatment ensured that the diamond surface was oxygen-terminated, thus ruling out the contribution to NV charge state conversion of a possible shift in the Fermi level due to chemically induced band bending. As schematically shown in **Figure 1b**, 30 keV $Ga^+$ focused ion beam (FIB) milling was performed at the outer endpoints of the micro-channels to expose them to the sample surface. The FIB milling was followed by the deposition of 70 nm thick Ag contacts through a stencil mask, on which electrical connections to the external circuit were subsequently wire-bonded. The final device consisted of two ~10 μm wide and ~100 μm long graphitic micro-electrodes spaced by a ~9 μm gap, as shown in **Figure 1c**.

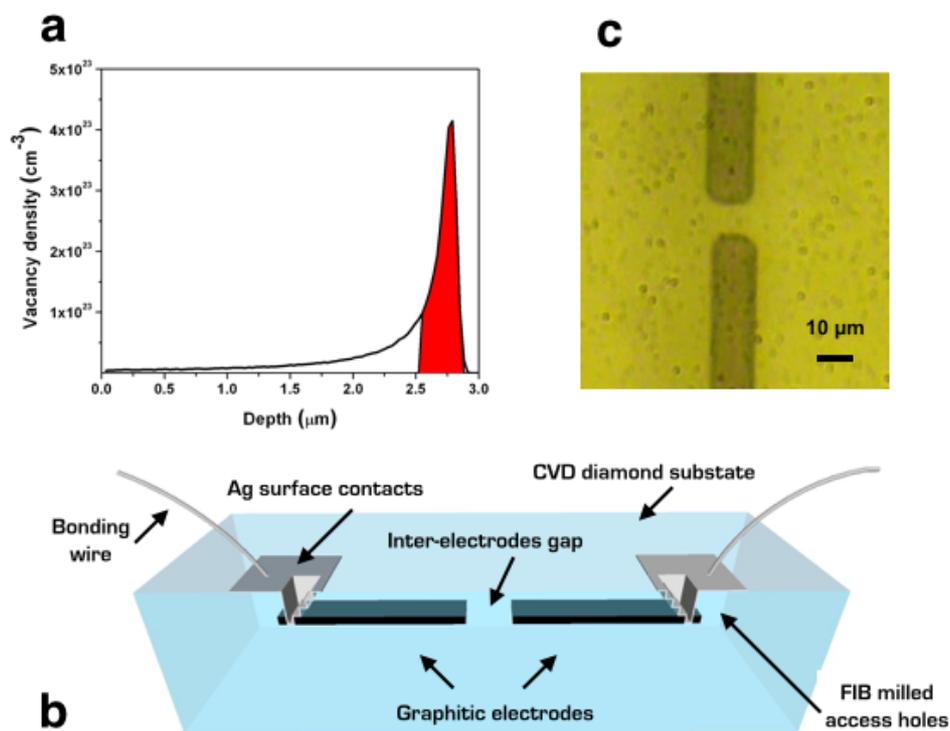

**Figure 1**: Fabrication of sub-superficial graphitic electrodes in diamond. **a)** Vacancy-density depth profile induced by 6 MeV $C^{3+}$ ions implanted at a fluence of $4 \times 10^{16}$ cm$^{-2}$, as evaluated by SRIM2013.00 Monte Carlo code [32]. The vacancy density exceeds the graphitization threshold in the region highlighted in red, where graphitic electrodes are formed upon thermal annealing. **b)** Schematic representation of the device: the electrical connections of the graphitic electrodes to the external circuit are fabricated by FIB milling at their outer endpoints, in order to expose them to a subsequent Ag contact deposition and wire-bonding. **c)** Transmission optical micrograph of the inter-electrode gap region of the fabricated device.

## 2.2. Electrical and optical characterization

Two-terminal current-voltage (I-V) characteristics were measured with a standard setup (Keitley 6487 picoammeter/voltage source), both in dark conditions and under laser illumination. Increasing voltage steps of 2 V and a constant source time of 300 ms were adopted, both in the acquisition of the I-V curves and of the photoluminescence (PL) spectra from the biased device. The injected current was measured during PL acquisitions to monitor the electrical stability of the device. PL measurements were performed on the inter-electrode gap region, under different bias conditions. The PL spectra were acquired with a Horiba Jobin Yvon HR800 Raman spectrometer equipped with a 1800

mm$^{-1}$ diffraction grating (∼0.1 nm spectral resolution). In all the experiments discussed in the following, photoexcitation was provided by a continuous 532 nm laser, with a power of 21.6 mW, as measured on the sample surface. A 20× air objective was employed to define a probed spot region of ∼3 μm both in diameter and focal depth, i.e. comparable with the penetration range of 6 MeV C$^{3+}$ ions in diamond, thus minimizing the effects of possible small stage drifts during the measurements at variable bias voltages. To account for slightly non-reproducible focusing conditions, all the reported PL spectra (as well as the combined PL/EL spectra reported in **Section 3.3**) were normalized to the intensity of the first-order Raman peak. Each PL acquisition from the biased device was followed by an acquisition at zero bias voltage, to both verify the absence of residual polarization effects and check the repeatability of the measurements.

## 3. RESULTS AND DISCUSSION

### 3.1 Electrical characterization

*I-V characteristic in dark conditions*

The I-V characteristic of the device in dark conditions is reported in **Figure 2a** (black line). The current measured in these experimental conditions will be referred to as "dark current" in the following. At increasing bias between 0 V and 250 V, a quasi-linear increase of the dark current was observed, indicating an ohmic diamond-graphite interface with a typical resistance value of 700 MΩ [22,33]. The I-V trend exhibited a moderate super-linear current increase at $V_{bias}$ > 250 V. As a critical bias voltage ∼300 V was reached, a sudden increase of the dark current was observed, with currents larger than 10 μA at 350 V. This behavior was interpreted in terms of Space-Charge-Limited Current (SCLC) caused by the injection of electrons at the ohmic contacts [34,35]. Accordingly to this model, the super-linear current increase observed at increasing voltage (i.e. in the 250−350 V range) is associated with the progressive filling of electron traps in the diamond band gap, while the abrupt current increase observed at $V_{bias}$ > 300 V indicates the complete filling of the available deep trap states. Assuming for simplicity a monoenergetical trap distribution in the diamond energy gap, the critical bias $V_c$ = 300 V represents the voltage at which the electron quasi-Fermi level overcomes the closest deep trap level $E_t$ to the conduction band [35]. While the assumption of an individual type of electron trap in the band gap is rather simplicistic, it is reasonable to assume that there is one predominant deep trap level which is responsible for the transition to the high-current-injection regime. The position of this level in the energy band gap was estimated according to the SCLC theory [35], as described below.

Firstly, the following equation expresses the higher threshold value $V_c$ = 300 V:

(1) $\qquad V_c = q\, N_t\, d^2 / (2\, \varepsilon_0\, \varepsilon_r)$

where $q$ is the elementary charge, $N_t$ is the trap density, $d$ = 9 μm is the width of the inter-electrode gap region and $\varepsilon_r$ = 5.5 is the dielectric permittivity of diamond. From Eq. (1), a $N_t = (2.2 \pm 0.3) \times 10^{15}$ cm$^{-3}$ value was derived. Considering the ∅~5 μm spot size of the 6 MeV C$^{3+}$ micro-beam adopted for the device fabrication, the implantation of stray ions with a fluence smaller than one order of magnitude in the inter-electrode region is expected [29]. Thus it is reasonable to ascribe the trap density to the residual concentration of ion-induced defects in the active region of the device after the thermal treatment.

Moreover, the following formula expresses the lower threshold voltage $V_{tr}$ = 250 V, where a deviation from the linear ohmic conduction becomes apparent [35]:

(2) $\qquad V_{tr} = 8/9 \cdot (q\, n_0\, d^2)/(\varepsilon_0\, \varepsilon_r\, \theta)$

where $n_0 = 2.3 \times 10^{10}$ cm$^{-3}$ is the free electron concentration at thermal equilibrium as estimated from the resistivity of the device at low voltages assuming an electron mobility of $\mu = 2200$ cm$^2$ V$^{-1}$ s$^{-1}$ [25], and $\theta$ is the ratio between free and trapped electrons when the electron quasi-Fermi level lies below the trap energy $E_t$ [35]. From Eq. (2), a value of $\theta = 7.5 \times 10^{-6}$, namely the fraction of the total charge density at the anode which is available for conduction at biases lower than $V_{tr}$ [34], was obtained.

Finally, the following formula expresses the $\theta$ value [34,35]:

(3) $\quad \theta = N_c/g/N_t \cdot \exp(-\Delta E_t/kT)$

where $N_c = 1 \times 10^{20}$ cm$^{-3}$ is the effective density of states in the conduction band, $g$ is the degeneracy of the trap level, $\Delta E_t$ is the difference between the energy corresponding to the bottom of the conduction band and the energy of the trap level, $k$ is the Boltzmann constant and $T$ is the temperature. From Eq. (3), a $\Delta E_t = 0.57$ eV was obtained under the assumption of $g = 2$ (see below). Indeed, such $\Delta E_t$ value is in good agreement with the position of the excited state of the NV$^-$ center, i.e. 0.6 eV from the bottom of the conduction band [36]. On the other hand, also the assumed $g = 2$ value is compatible with the excited state of the NV$^-$ center [37]. Thus, we conclude that the electron quasi-Fermi energy at the critical threshold $V_c$ overcomes the excited state of the NV$^-$ center, so that all NV$^-$ centers (and all deeper levels in the band gap) are filled. Consequently, the SCLC model suggests that an almost complete conversion by electron capture from the neutral to the negative charge state of the NV center is achieved at $V_{bias}=V_c$ (see below for a more detailed discussion).

The I-V trend in high-current-injection regime (i.e., at $V_{bias} > 400$ V) is suitably described with a Poole-Frenkel (PF) model [38], as demonstrated by the linearization of the "asinh($i/V$) vs $V^{1/2}$" trend (see **Figure 2b**) deriving from the characteristic PF expression $I \propto V \sinh(aV^{1/2}/kT)$ [39], in good agreement with previous findings on similar diamond devices [30]. According to this model, the conduction mechanism is dominated by the thermally induced electron emission from traps into the conduction band under the effect of the strong applied electric field [35]. Concurrently to the transition to this high-current-injection regime, an intense electroluminescence (EL) emission was observed from the inter-electrode gap region. More details on this effect and the above-mentioned interpretation are reported in **Section 3.3**.

*I-V characteristics under laser illumination*

I-V curves were also acquired under laser illumination of the inter-electrode gap region of the device (**Figure 2**, red line) with the same experimental parameters. The current measured in these conditions will be referred to as "total current", as it consists of a further contribution ("photocurrent" in the following) arising from photo-generated carriers. A quasi-linear trend in the I-V curve was observed at low bias voltages (<100 V), while a super-linear increase of the total current was apparent at higher voltages. At bias voltages smaller than $V_c$, the ∼50% current increase under laser illumination indicates that the optical excitation promotes the de-trapping of carriers.

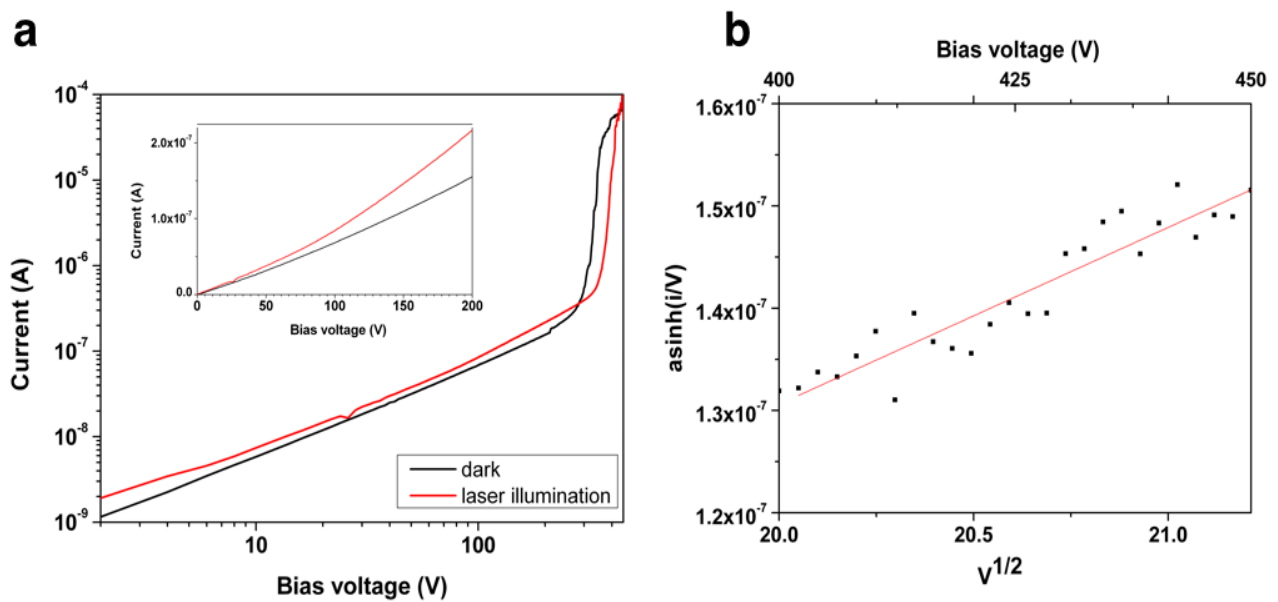

**Figure 2**: **a)** I-V characteristic in dark conditions (black line) and under laser illumination (red line) of the inter-electrode gap region. In the inset the data are plotted in linear scale in the 0-200 V range for sake of readability. **b)** PF electrical conduction in high-current-injection regime. On the basis of the PF formula $I \propto V \sinh(aV^{1/2}/kT)$, the data in the 400-450V range are linearized by suitably re-scaling the quantities reported on the axes.

Moreover, upon laser illumination the transition to a high-current-injection regime was shifted at a higher threshold voltage (~350 V). This evidence indicates that the optically-induced detrapping of the carriers represents a competing process with respect to the electron trap filling in the interelectrode gap, thus supporting the above-mentioned interpretation of the transition between low- and high-current-injection regimes based on the SCLC process.

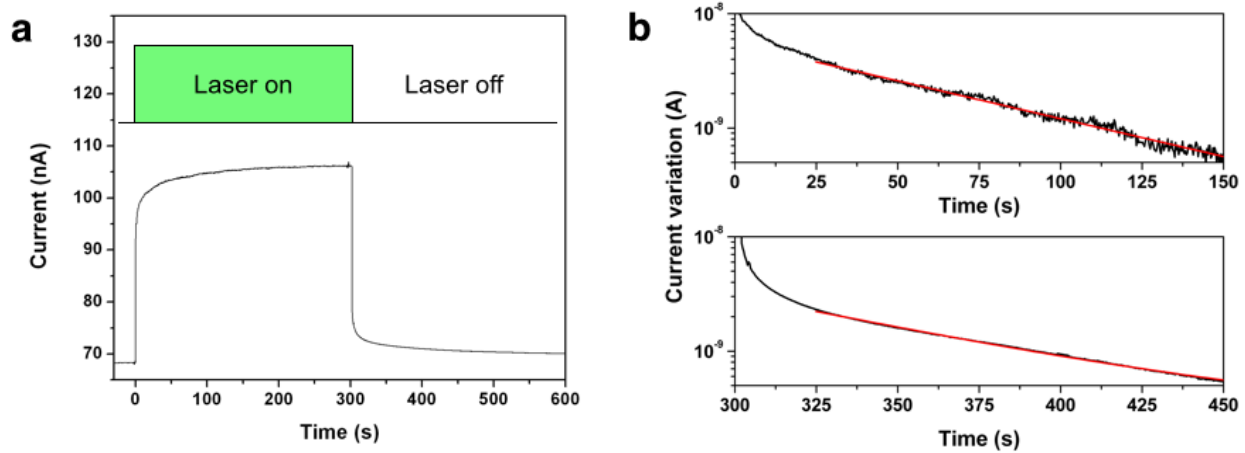

**Figure 3**: **a)** Total current chronogram under variable laser illumination of the inter-electrode gap region, at an applied bias voltage of 100 V. As schematically shown in the inset, the laser illumination is turned on at time $t = 0$ s and switched off after 300 s. **b)** Chronograms of the absolute value of the difference between the measured current and the asymptotic baseline (black lines) in correspondence of the laser turn-on (top) and subsequent switch-off (bottom) together with the corresponding exponential fits (red lines).

The carriers trapping/detrapping mechanisms were further investigated with current measurements under variable laser illumination. As shown in the chronogram reported in **Figure 3a**, the current in the device biased at 100 V was sampled in concurrence with the turning on and the switching off of the laser illumination, to investigate the related characteristic trapping/detrapping times.

As shown in **Figure 3b**, both the transitions were characterized by the sum of a fast exponential term and a slower one. In both cases, the former term was associated with a characteristic time smaller than the sampling period (i.e. 300 ms). We ascribed this first component to the trapping/detrapping of shallower defects. The slower components of the turn-on and switch-off current transients were characterized by significantly longer decay times, i.e. $(65.6 \pm 0.4)$ s and $(69.8 \pm 0.7)$ s, respectively. We ascribed this second component to the trapping/detrapping of deep defects in the band gap. These trapping/detrapping effects can be associated with the photo-induced conversion of the charge state of the probed NV centers [14,40], although the possible contribution of other deep levels associated with different defects cannot be ruled out. Temperature-dependent photocurrent measurements could indeed be useful to further elucidate this issue. It is however worth noting that the measured characteristic PL decay times are significantly shorter than what recently reported for trapping/detrapping of the P1 center (i.e. substitutional nitrogen defects) in photoconducting diamond [41], so that at least this possible attribution can be ruled out.

**3.2 Charge state conversion**

The PL emission from the inter-electrode gap region of the device was measured under different applied bias voltages, with the purpose of assessing the variation of the photophysical properties of the probed NV centers as a function of the injected current between the electrodes. In **Figure 4a** the PL spectrum from the unbiased device is reported together with PL spectra acquired at increasing bias voltages in the 0−350 V range, i.e. below the transition to the high-current-injection regime (and the concurrent EL emission) is observed. The spectra exhibit the typical PL features of an "optical grade" diamond sample after low-fluence ion irradiation and subsequent annealing, such as the intense first-order Raman peak at 572 nm (corresponding to a 1332 cm$^{-1}$ Raman shift), and the NV$^0$ and NV⁻ zero phonon lines (ZPLs) at 575 nm and 638 nm, respectively, together with their associated phonon sidebands at higher wavelengths [8]. Additionally, the luminescence peak at 740 nm is attributed to residual GR1 centers, i.e. to neutral isolated vacancies [42]. A systematic variation of the relative intensities of the PL features associated with the different NV charge states is clearly visible. For sake of readability, in **Figure 4b** we report the "relative" PL spectra acquired from the biased device, i.e. after the subtraction of the reference PL spectrum acquired at zero bias. The most apparent feature of the data displayed in **Figure 4b** is a substantial decrease in the $NV^0$ emission at increasing bias, concurrently to an increase in the NV⁻ emission. It is worth noting that the apparent decrease in the NV⁻ ZPL emission is due to the large decrease of the background represented by the $NV^0$ phonon sideband, which results in a lowering of the NV⁻ ZPL peak baseline. If the above-mentioned background is subtracted, the intensity of the NV⁻ ZPL actually increases with the applied bias, as indicated by the increasing local maximum observed at $\lambda = 638$ nm, and confirmed by the visible increase in intensity of the NV⁻ phonon sideband at higher wavelengths. It is also worth noting that, similarly to the $NV^0$ emission, the GR1 luminescence peak decreases at increasing biases, as indicated by the deepening dip at 740 nm.

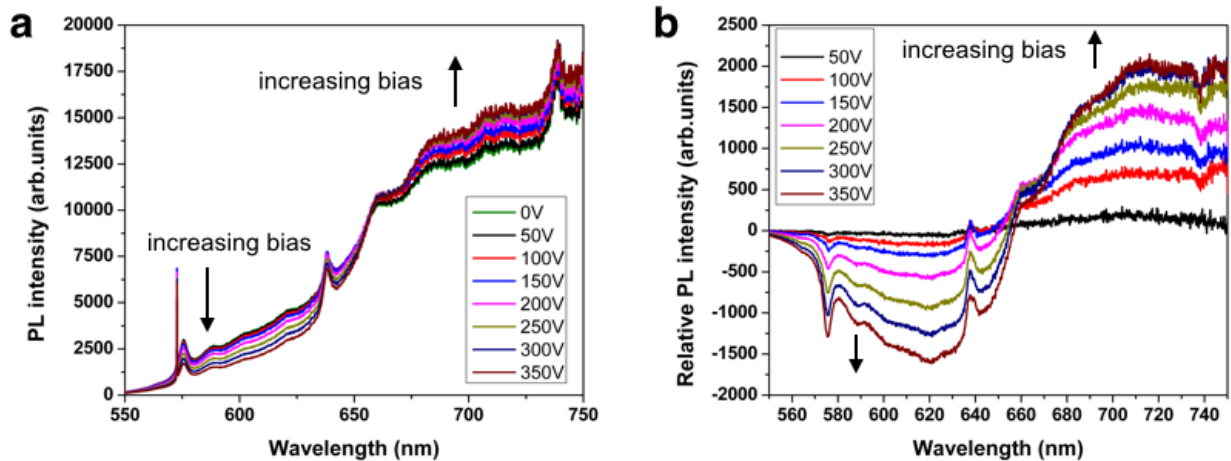

**Figure 4**: **a)** PL spectra acquired from the inter-electrode gap region at increasing applied bias voltages. **b)** Relative PL spectra acquired from the biased sample, after subtraction of the reference PL spectrum acquired from the unbiased device.

The current-dependent values of the integrals $A_0(i)$ and $A_-(i)$ calculated respectively in the 572-580 nm ($NV^0$ ZPL) and 634-642 nm ($NV^-$ ZPL) spectral intervals were evaluated after the subtraction of their respective baselines. The $A_0(i)$ and $A_-(i)$ values were then normalized to their respective reference values acquired from the unbiased device, and the resulting $A_0(i)/A_0(0)$ and $A_-(i)/A_-(0)$ normalized values are reported in **Figure 5a** as a function of the corresponding injected current $i$. A relative uncertainty of 5% on the ZPL integrals was estimated.

The data reported in **Figure 5a** exhibit increasing and decreasing linear trends, respectively for the $NV^-$ (black square dots) and $NV^0$ (red circular dots) normalized ZPL emissions. A quantitative evaluation of the relative concentrations of NV centers in different charge states was performed under the assumption that the current-dependent local concentration of $NV^{0/-}$ centers in the probed region, i.e. $n_{0/-}(i)$, is directly proportional to the correspondent ZPL intensity $A_{0/-}(i)$. Thus it follows that:

(4) $\quad n_{0/-}(i) / n_{0/-}(0) = A_{0/-}(i) / A_{0/-}(0)$

where the the "$0/-$" subscript refers alternatively to either of the two NV charge states.

Furthermore, we assumed that the probed NV centers only change their charge states between the $NV^0$ and $NV^-$ ones, i.e. that a possible (and yet to be conclusively demonstrated) population fraction of "dark" $NV^+$ centers remains (if at all present) unchanged under various injection conditions. Under these assumptions, it follows that the sum of the overall population of $NV^-$ and $NV^0$ centers ($n_-(i)$ and $n_0(i)$, respectively), is constant for any current flowing in the device:

(5) $\quad n_-(i)+n_0(i)=n_{tot}$

From the linear fitting in the 0−405 nA range of the "$n_{0/-}(i)/n_{0/-}(0)$ vs $i$" data reported in **Figure 5a** it is possible to express each NV charge state concentration as follows:

(6) $\quad n_{0/-}(i) = n_{0/-}(0) \cdot (1+k_{0/-} \cdot i)$

where $k_0 = (-0.99 \pm 0.03)$ µA$^{-1}$ and $k_- = (+1.06 \pm 0.06)$ µA$^{-1}$ are the respective angular coefficients. From Eqs. (5) and (6) it is possible to obtain, for the unbiased device:

(7) $\quad n_-(0) / n_0(0) = |k_0| / k_-$

It follows that $n_-(0) / n_{tot} = (0.52 \pm 0.06)$ and $n_0(0) / n_{tot} = (0.48 \pm 0.06)$. From these values and the previously derived linear coefficients it is possible to obtain the relative concentrations of the different NV charge states as a function of the injected current (see **Figure 5b**):

(8) $\quad n_0(i) / n_{tot} = (0.48 \pm 0.06) - (0.48 \pm 0.07) \cdot i\ [\mu A]$

$\quad\quad n_-(i) / n_{tot} = (0.52 \pm 0.06) + (0.55 \pm 0.08) \cdot i\ [\mu A]$

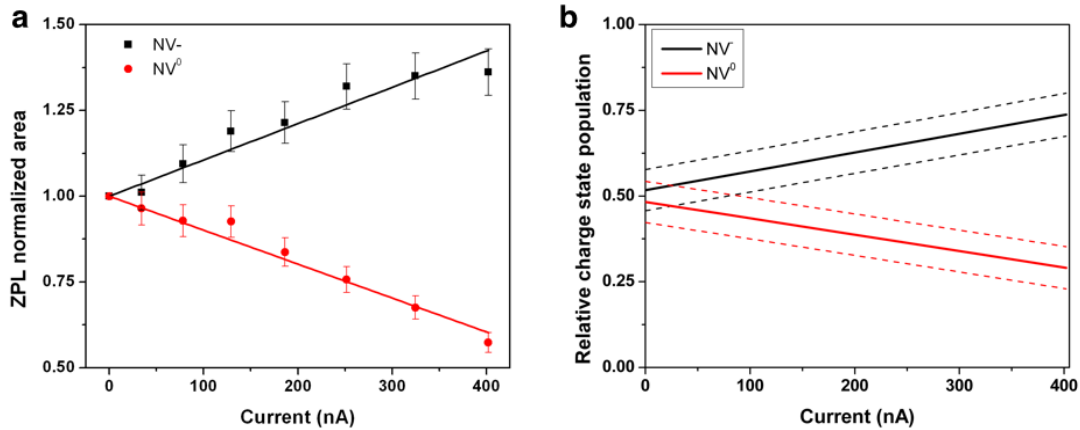

**Figure 5**: **a)** Normalized ZPL intensity from NV$^-$ ($A_-(i)/A_-(0)$, black square dots) and NV$^0$ ($A_0(i)/A_0(0)$, red circular dots) centers as a function of the total injected current $i$. Linear fitting curves of the data in the 0-405 nA range are reported by the continuous lines. **b)** Variation of $n_-(i)/n_{tot}$ (black line) and $n_0(i)/n_{tot}$ (red line) as a function of the total injected current $i$, as resulting from **Eq. (8)**. The dashed lines indicate the uncertainty associated to our estimation, as derived from the fitting procedure.

It is worth noting that a relative concentration of NV$^-$ centers of $n_-/n_{tot} = (0.74 \pm 0.09)$ is achieved in correspondence of an injected current of ∼400 nA, corresponding to a ~40% increase with respect to the unbiased device. If it is considered that the steady-state NV$^-$ relative population is always smaller than 0.75 under non-resonant excitation [36], our results indicate that the device achieves the maximum possible conversion to the negative charge state under electrical control. It is worth remarking that this maximum $n_-/n_{tot}$ value is achieved at a bias voltage (350 V) slightly below the voltage corresponding to the transition to the high-current-injection regime (red curve in **Figure 2**). This evidence represents a further confirmation of the SCLC model elucidated in **Section 3.1**, which is based on the progressive electron trapping of the NV centers, i.e. to their progressive conversion to a negative charge state, until the transition to a high-current-injection regime occurs when the trap levels are filled. Such interpretation is further supported by the observed decrease in the GR1 PL emission associated with the neutral vacancy, which could be similarly ascribed to its conversion upon electron trapping to the negative charge state, not optically active under 532 nm laser excitation [42]. On the other hand, an alternative explanation of the change of the NV charge state based on a variation in the surface electric potential can be discarded since a bulk region is probed by means of sub-superficial electrodes. In addition, the fact that the sample surface was oxidated during the fabrication process allows to rule out a significant shift of the Fermi level due to a chemically/electrically induced band bending associated with the charge state conversion of superficial NV centers [17] within the focal depth of the air objective. It is also worth noting that, since the injection of a hole current has been demonstrated to promote the conversion of the NV centers population to the neutral charge state [18], electron trapping/detrapping from/to the conduction band is identified as the predominant process determining the NV charge state in low-current-injection regime, with respect to hole trapping/detrapping. This interpretation based on a unipolar charge transport is supported by previous independent experimental results obtained from "Ion Beam Induced Charge" (IBIC) measurements, that demonstrated an electron-dominated contribution to the charge transport in the same type of diamond substrates fabricated with sub-superficial graphitic electrodes [43].

The proposed trapping/detrapping processes associated with electrically-induced NV charge state conversion are: the capture of an electron by an $NV^0$ center resulting in the formation of an $NV^-$ state [44] and, conversely, the detrapping of one electron from the $NV^-$ center to the conduction band, upon either electrical or optical excitation, leading to a switching to the neutral $NV^0$ state [18,36,40]:

(9)  a) $NV^{0*} + e \rightarrow NV^-$

   b) $NV^{-*} \rightarrow NV^{0*} + e$

Consistently with what reported in [14,40], we assume that the electron trapping/detrapping mainly occurs at the excited states (labeled as $NV^{-*}$ and $NV^{0*}$ in **Eq. 9**) of the two types of center, due to their closer proximity to the conduction band edge with respect to their respective ground states. In agreement with what discussed in previous works [11,16], the relative increase in the $NV^-$ concentration originates from a shift in the electron quasi-Fermi level, due to the progressive filling of trap states in the band-gap at increasing currents. As highlighted by the electrical characterization discussed in **Sect. 3.1**, the transition to a high-current regime occurs when the electron quasi-Fermi level reaches the relevant trap state placed at 0.57 eV below the conduction band. Such transition indicates that a complete filling of all the deeper level in the band gap is achieved, including the conversion of the NV centers to the negative charge state by electron capture.

A schematic representation of the concurring electron trapping/detrapping phenomena responsible for the NV charge state modification in low-current-injection regime is reported in **Figure 6**. In the figure, the green band represents an excitation energy corresponding to the 532 nm laser from the conduction band. The diagram represents the energy levels corresponding to the ground and excited states of the NV$^-$ and NV$^0$ centers [8]. The ground states of the NV$^-$ and NV$^0$ centers are located at 2.54 eV and 4.3 eV below the conduction band, respectively [16,36]. Similarly, the corresponding excited states NV$^-$* and NV$^0$* lay at 0.6 eV and 2.14 eV below the conduction band, respectively [16]. The ground and excited state of the GR1 center (respectively 2.86 eV and 1.18 eV below the conduction band), i.e. the neutral single vacancy V$^0$ are also reported [37]. The negatively charged vacancy V$^-$ (usually referred as ND1 centre) has a ground state placed at 2.6 eV below the conduction band; its excited state (3.15 eV above the ground state) lays within the conduction band [42]. The electronic transitions involved in the charge state modification of the afore-mentioned centers can be thus summarized as follows. After carrier injection through the ohmic diamond/graphite interface, the electrical conduction in the inter-electrode gap region occurs within the conduction band concurrently with trapping/detrapping phenomena (labeled as "1-4" and "8,10" in **Figure 6**). The trapping/detrapping processes can involve either shallow traps laying at energy $E_{sh}$ (labeled as "1" and "2" in **Figure 6**), or the excited state of NV$^-$ centers (labeled as "3" and "4" in **Figure 6**) as described by **Eq. (9)**. In the latter case, the electron de-trapping from the NV$^-$ center results in its conversion to the NV$^0$ state [15] (labeled as the blue dashed arrows "5" in **Figure 6**). Conversely, the transition of the NV$^0$ to the negative charge state NV$^-$ (labeled as the green dashed arrows "6" in **Figure 6**) is assumed to occur when an electron from the conduction band is trapped at the available energy level at 0.57 eV, corresponding to the NV$^-$ excited state (transition labeled as "4" in **Figure 6**). At increasing electron currents, the progressive filling of the trap states available in the band gap results in an increase in the electron quasi-Fermi level $E_{Fn}(i)$ (labeled as the green dashed arrow "A" in **Figure 6**), thus stabilizing the negative charge state of the NV center. Conversely, at decreasing injected currents the electron quasi-Fermi level decreases upon electron detrapping processes (labeled as the blue dashed arrow "B" in **Figure 6**). Similarly to what described for the NV centers, the decrease in the PL intensity of the GR1 center at increasing currents is caused by the conversion to the ND1 negative charge state (labeled as the green dashed line "7" in **Figure 6**), which in turn is a consequence of the non-radiative electron trapping from the band gap to the V$^-$ ground state (green line "8" in **Figure 6**). The inverse charge state conversion is presented for completeness (dashed blue line "9" in **Figure 6**) as the result of an electron transition from the ND1 ground state to the conduction band, where its excited state is laying (labeled as the blue arrow "10" in **Figure 6**).

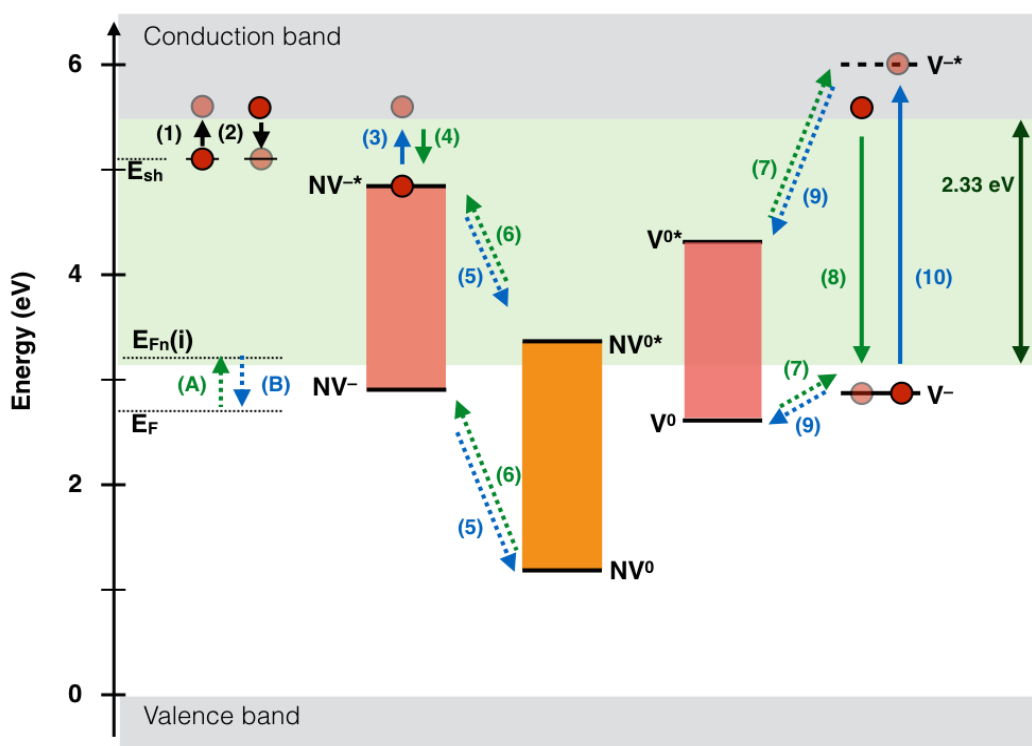

**Figure 6**: The region highlighted in green indicates the energy levels, which can be optically detrapped upon 532 nm laser excitation (2.33 eV). Schematic representation (in scale) of the electron trapping/detrapping

mechanisms occurring in the inter-electrode gap region. Detrapping (labeled as "1") and trapping (labeled as "2") from/to shallow trap levels at a generic energy $E_{sh}$ contribute to the measured photocurrent (see **Figure 3**). Electron detrapping (blue line, labeled as "3") from an excited NV⁻ center results in the conversion to its neutral charge state (blue dashed line, labeled as "5"). Conversely, electron trapping at the NV⁻* level (green line, labeled as "4") causes the conversion of the NV⁰ center to its negative charge state (green dashed line, labeled as "6"). Similarly, an electron trapping to (green arrow, label "8") or detrapping from (blue arrow, label "10") the ground state of the negative vacancy results respectively in the V⁰ → V⁻ (blue dashed line, labeled as "7") or V⁻ → V⁰ conversion (green dashed line, labeled as "9"), respectively.

**3.3 Electroluminescence**

The study of the electrically controlled charge state conversion was limited by the transition to a high-current-injection regime observed at bias voltages higher than 350 V (see **Figure 3a**). This transition was fully reproducible and stable over multiple excitation cycles. A non-negligible electrically induced luminescence was detectable for injected currents larger than 10 µA, i.e. above the transition to the high-current-injection regime, while no EL emission could be detected at lower currents within the experimental sensitivity.

EL was investigated at increasing bias voltages under no laser illumination. An optical micrograph acquired at a bias voltage of 500 V is shown in **Figure 7a**, exhibiting an intense red emission visible by naked eye in the inter-electrode gap region. EL spectra acquired at 350 V and 500 V bias voltage with the same light collection setup adopted for PL measurements are reported in **Figures 7b** and **7c**, respectively. The main spectral feature is the emission from the NV⁰ center, which is clearly characterized by the ZPL at 575 nm and its phonon sidebands at higher wavelengths. The NV⁻ emission component is absent from the EL spectrum, consistently with previous reports of the absence of such emission under electrical excitation [19,42]. Moreover, two sharp EL emission lines (not active under optical excitation at 532 nm), are clearly visible at $\lambda = 563.5$ nm and $\lambda = 580$ nm.

As shown in **Figures 7b** and **7c**, at increasing injected currents the two above-mentioned EL lines display a similar increase in emission intensity, which is more pronounced with respect to the $NV^0$ emission. Indeed, the $NV^0$ ZPL line is barely distinguishable at 500 V bias (see **Figure 7c**). The strong correlation between the emission intensities of the $\lambda = 563.5$ nm and $\lambda = 580$ nm lines suggests a possible attribution to the same defect family or to different charge states of the same complex. Although an exhaustive investigation of these previously unreported EL emission lines goes beyond the scope of the present work, they are tentatively attributed to self-interstitial defects, i.e. pairs of nearby interstitial atoms along the <100> axis, consistently with what was previously observed in cathodoluminescence and photoluminescence under 488 nm laser excitation, in ion- and electron-irradiated diamonds [42,45].

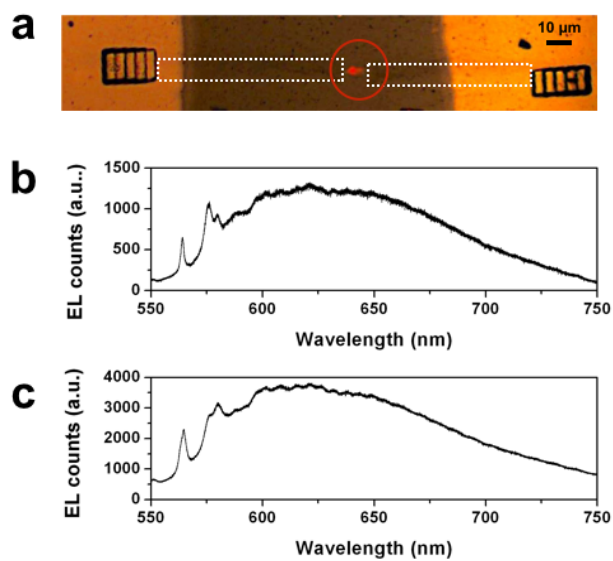

**Figure 7**: **a)** Reflection optical micrograph of the device biased at a voltage of 500 V. The edges of the (otherwise barely visible) graphitic electrodes are highlighted by the white dashed line for sake of clarity. An intense EL emission (circled in red) is clearly visible as the red spot in the inter-electrode gap region. **b)** EL spectrum acquired from the inter-electrode gap region at a bias voltage of 350 V. **c)** EL spectrum acquired from the same region at a bias voltage of 500 V.

A quantitative analysis of the EL emission intensity as a function of the injected current was limited by the strong spectral overlap of the $NV^0$ ZPL with the 580 nm emission line. Furthermore, any indication on the NV charge state conversion was prevented by the absence of $NV^-$ emission in EL regime. However, a qualitative analysis of the process was possible by a PL investigation of the device in high-current-injection regime, i.e. at bias voltages higher than 400 V under laser illumination. In this case, the luminescence spectra acquired in the inter-electrode gap region included both a PL and an EL contribution. The former contribution was associated to the optical excitation of $NV^0$, $NV^-$ and GR1 centers, while the latter one consisted in the $NV^0$ spectral component and the two afore-mentioned interstitial-related emissions. **Figure 8a** shows such combined "PL+EL" spectra acquired at increasing bias voltages (400 V, 450 V and 500 V), while also reporting for comparison a PL-only emission spectrum acquired from the unbiased device.

At increasing bias voltages, the PL+EL spectra exhibit an increase in the emission intensity over the whole spectral range. The corresponding injected current values can be found in **Figure 2a** (red curve). Similarly to what reported in **Figure 4b**, for sake of readability **Figure 8b** shows the "relative" PL+EL spectra obtained after subtraction of the reference spectrum collected from the unbiased device. Firstly, the significant increase in emission intensity from the interstitial-related emission lines at 563.5 nm and 580 nm at increasing bias voltages is clearly visible. On the other hand, as previously mentioned the increase in the $NV^0$ emission is significantly less pronounced with respect to the two afore-mentioned peaks. Nevertheless, a non-negligible increase in the $NV^0$ emission can be appreciated by considering the increase in its phonon sidebands. While the overall increase in $NV^0$ emission at increasing injected currents arises in large part from the progressive increase of EL emission (see **Figures 7b** and **7c**), it cannot be ruled out that the $NV^{-*} \rightarrow NV^{0*} + e$ conversion process is also favored by the underlying electrical conduction mechanism. On the other hand, the deepening local minimum at $\lambda = 638$ nm indicates that the intensity of the $NV^-$ emission (and thus the $NV^-$ centers concentration) decreases at increasing bias voltages in the high-current-injection regi-me. Remarkably, this trend is opposite to what was observed at bias voltages below the critical threshold of 350 V, i.e. a progressive increase in the NV⁻ charge state concentration at increasing voltages (see **Figure 4b**). It is interesting to notice that such behavior is the same as reported in the study of the charge-state conversion process of NV centers in p-i-n devices in forward bias [20]. In that case, however, the charge state conversion at biases smaller than the built-in voltage was interpreted in terms of a pure bending of the energy bands.

Finally, it is worth remarking that a significant decrease in the $V^0$ emission is still observed at increasing biases, as confirmed by the deepening local minimum observed at $\lambda = 740$ nm. Differently from what observed for the NV⁻ emission, this effect is in line with the same trend observed in low current-injection regime (see **Figure 4b**).

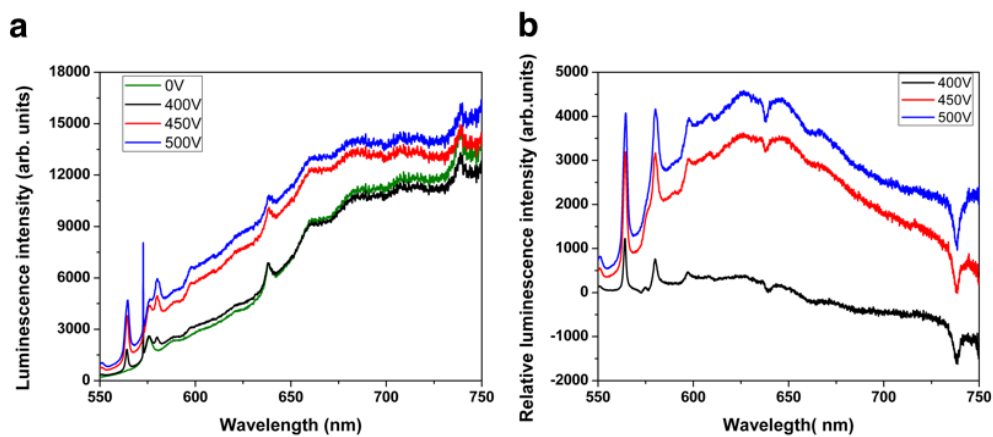

**Figure 8**: **a)** PL+EL spectra acquired under laser excitation at increasing bias voltage in the high-current-injection regime ($V_{bias}$ = 400-500 V). The PL spectrum acquired from the unbiased device (green line) is also reported for comparison. **b)** Corresponding relative luminescence spectra obtained after subtraction of the reference PL spectrum acquired at zero bias.

The observation of EL in a high-current-injection regime dominated by a Poole-Frenkel conduction is somewhat surprising, since the EL emission was previously demonstrated in p-i-n devices to be correlated with electron-hole recombination processes at NV sites, originating from double carriers injection at the electrodes [19,46]. While a hole injection in low-current regime could not be ruled out in first instance, the results discussed in **Section 3.2**, combined with the previous findings on NV conversion to the neutral charge state upon hole capture [18], indicate a net contribution to the modification of charge state of the nitrogen-vacancy center which is ascribable to a majority electron current [18], an interpretation which is fully consistent with previous independent experimental results obtained from "Ion Beam Induced Charge" (IBIC) measurements in similar devices [43]. On the other hand, the increasing conversion to the $NV^0$ charge state at increasing bias suggests that the transition to a high-current regime involves the injection of a significantly higher density of holes in the inter-electrodes gap, with respect to the low-current behavior. This interpretation is consistent with previous reports [19,46] on the observation of EL emission in presence of a non-negligible hole component in the high current-injection regime. Finally, it is worth mentioning that, on the basis of the present results, an interpretation of the obtained results in high-current regime exclusively based on a thermally-stimulated trapping/detrapping process consistent with the Poole-Frenkel conduction mechanism cannot be ruled out. In this latter interpretation, a high rate detrapping, combined with a high pump electron current can in principle fulfill the same role as the electron-hole injection and recombination in the electrical stimulation of the $NV^0$ emission.

## 4. SUMMARY AND CONCLUSIONS

In this work, we demonstrated that sub-superficial graphitic electrodes can effectively be exploited to control the charge state of nitrogen-vacancy complexes in diamond. The electrical characterization of the device highlighted an ohmic conduction mechanism in the inter-electrodes gap at low voltages; at higher biases (>350 V), a significant current increase was observed and was ascribed, according to the SCLC model, to the filling of traps in the inter-electrodes gap. Such model enabled us to estimate the position of the dominant trap level at 0.57 eV below the conduction band. This value is compatible with the position of the excited state of the negatively-charged NV center.

In the low-current-injection regime a linear increase in the NV⁻ charge state population at the expense of the neutral charge state NV⁰ was observed by ensemble PL spectral measurements at increasing injected currents. Consistently with the above-mentioned SCLC model, this process was interpreted as the result of trapping of electrons injected in the active region of the device. The electrical control of the NV charge state enabled to reach a ∼75% fraction of negatively charged NV centers upon the injection of a stationary 400 nA current.

The high current-injection regime was associated with an intense EL emission. Differently from devices fabricated by He⁺ ion irradiation [29], the implantation of $C^{3+}$ ions prevented the formation of additional color centers related to foreign impurities in the diamond lattice. Interestingly, previously unreported sharp EL emission peaks at 563.5 nm and 580 nm were observed, exhibiting a stronger dependence from the injected current with respect to the NV⁰ center. These emissions were tentatively attributed to interstitial defects, consistently with previous CL studies [42,45]. Finally, the PL contribution of NV⁻ centers in high current-injection regime indicated a decrease in the negatively-charged concentration of the NV defect, consistently with an increased hole current injection in the device or, alternatively, with a high-rate thermally-stimulated electron detrapping process associated with the Poole-Frenkel conduction mechanism.

Apart from shedding light on the fundamental mechanisms underlying charge state modifications of NV centers in diamond, the results reported in this work prospect promising applications, which can be achieved by decreasing the electrodes distance by means of already demonstrated advanced ion-beam-lithography techniques exploiting implantation masks with sub-micrometric resolution [27]. This would offer the significant advantage of achieving high state conversion efficiency while significantly reducing the operating voltage. In perspective, these developments could contribute to stabilize the charge state of deep NV centers preventing an uncontrolled blinking. The technique can thus enable the preparation and control of the charge state of deep NV centers (which are kno-wn to be characterized by better spin coherence properties with respect to superficial defects), thus leading to promising applications in quantum computing and quantum sensing.

## ACKNOWLEDGEMENTS

This research was supported under the following schemes: "DIESIS" project funded by the Italian National Institute of Nuclear Physics (INFN) - CSN5 within the "Young research grant" scheme; "FIRB Future in Research 2010" project (CUP code: D11J11000450001) funded by the Italian Ministry for Teaching, University and Research (MIUR); "A.Di.N-Tech." project (CUP code: D15E13000130003) funded by the University of Torino and Compagnia di San Paolo in the framework of the "Progetti di ricerca di Ateneo 2012" scheme; "Diamond Microfabrication" experiment at the "Nanofacility Piemonte" laboratory of INRiM.